\documentclass[
	groupedaddress,
	amsmath,amssymb,
	prb,
	floatfix,
	superscriptaddress,
  twocolumn,
	aps
 ]{revtex4-2}

\usepackage{graphicx}%
\usepackage{dcolumn}%
\usepackage{bm}%
\usepackage[mathlines]{lineno}%
\usepackage{xcolor}
\usepackage{soul}
\usepackage{adjustbox}
\usepackage{nth}
\usepackage{hyperref}
\usepackage{cleveref}
\usepackage[T1]{fontenc}
\usepackage{nth}
\usepackage{amsmath}
\usepackage{xr}
\begin{document}

\title{Modeling the Equilibrium Vacancy Concentration in Multi-Principal Element Alloys from First-Principles}
\author{Damien K. J. Lee}
\affiliation{Laboratory of materials design and simulation (MADES), Institute of Materials, \'{E}cole Polytechnique F\'{e}d\'{e}rale de Lausanne}
\author{Yann L. M{\"u}ller}
\affiliation{Laboratory of materials design and simulation (MADES), Institute of Materials, \'{E}cole Polytechnique F\'{e}d\'{e}rale de Lausanne}
\author{Anirudh Raju Natarajan}
\email{anirudh.natarajan@epfl.ch}
\affiliation{Laboratory of materials design and simulation (MADES), Institute of Materials, \'{E}cole Polytechnique F\'{e}d\'{e}rale de Lausanne}
\affiliation{National Centre for Computational Design and Discovery of Novel Materials (MARVEL), \'{E}cole Polytechnique F\'{e}d\'{e}rale de Lausanne}

\begin{abstract}
Multi-principal element alloys (MPEAs), also known as high-entropy alloys, have garnered significant interest across many applications due to their exceptional properties. Equilibrium vacancy concentrations in MPEAs influence diffusion and microstructural stability in these alloys. However, computing vacancy concentrations from \emph{ab-initio} methods is computationally challenging due to the vast compositional space of MPEAs and the complexity of the local environment around each vacancy. In this work, we present an efficient approach to connect electronic structure calculations to equilibrium vacancy concentrations in MPEAs through embedded cluster expansions (eCE) and rigorous statistical mechanics methods. Using first-principles calculations and Monte Carlo simulations informed by eCE, we assess the variation in vacancy formation with alloy composition and temperature. Our method is demonstrated on a nine-component MPEA comprised of elements in groups 4, 5, and 6 of the periodic table. Correlations between alloy chemistry, short-range order, and equilibrium vacancy concentrations in alloys containing up to 9 different elements are analyzed. The vacancy concentration of refractory alloys increases with the addition of group 4 elements or elements whose mixing is energetically unfavorable. The insights into vacancy behavior and the efficient computational framework presented in this study serve as a guide for the design of complex concentrated alloys with controlled vacancy concentrations.
\end{abstract}

\maketitle

\section{Introduction}

Defects in crystalline materials play a crucial role in governing the rate of microstructural evolution and the properties of multicomponent alloys. For instance, substitutional diffusion in metallic alloys is mediated by vacancies \cite{balluffi_introduction_2005}. Although most alloys do not have an energetic preference for the formation of vacancies, a small concentration of these point defects nevertheless exists at equilibrium in most crystalline materials due to entropic effects. As diffusion constants are (to first order) proportional to the equilibrium vacancy concentration \cite{van_der_ven_vacancy_2010}, an accurate prediction of equilibrium vacancy compositions is essential to quantify multicomponent transport coefficients.

The presence of chemically diverse local environments in multicomponent alloys could result in a complex defect energy landscape \cite{wang_thermodynamics_2017,goiri_role_2019,behara_role_2024,li_vacancy_2024}. Vacancies are expected to preferentially segregate to environments with low vacancy formation energies. As a result, short-range order is expected to play a significant role in determining the vacancy concentrations of compositionally complex alloys. First-principles calculations coupled with statistical mechanics calculations have provided critical insights into the role of short-range order on equilibrium vacancy concentrations in binary alloys \cite{behara_fundamental_2024,behara_role_2024,thomas_thermodynamic_2024,belak_effect_2015, goiri_role_2019}. However, correlations between alloy chemistry, short-range order, and equilibrium vacancy concentrations in alloys containing three or more elements remain elusive.

Multi-principal element alloys (MPEAs), also known as high-entropy alloys, are attractive candidates for energy storage, catalytic, and structural applications \cite{yeh_nanostructured_2004, yeh_formation_2004, cantor_microstructural_2004, miracle_critical_2017, schweidler_high-entropy_2024, rost_entropy-stabilized_2015, oses_high-entropy_2020, ouyang_rise_2024,abu-odeh_simple_2024}. Several fundamental questions \cite{divinski_mystery_2018, tsai_sluggish_2013} about transport mechanisms in MPEAs need to be addressed to enable designers to tune these materials for commercial applications. In fact, very little is known about finite-temperature equilibrium vacancy concentrations in multicomponent alloys, a key ingredient for computing kinetic coefficients. Previous studies have employed density functional theory (DFT) calculations to compute zero Kelvin defect energetics of vacancies \cite{zhang_ab_2022, li_first_2019, mizuno_defect_2019, guan_chemical_2020} in multicomponent alloys. These defect energies are then used within finite-temperature approximations to estimate vacancy concentrations at elevated temperatures. The computational cost of DFT limits the size of simulation cells and the number of defect energies that can be computed. As a result, techniques that directly use DFT simulations to compute vacancy concentrations often sample only a small portion of the configurational and compositional space spanned by MPEAs.

MPEAs composed of elements from groups 4, 5, and 6 of the periodic table are an emerging class of materials for high-temperature structural applications \cite{natarajan_crystallography_2020,george_high-entropy_2019,muller_first-principles_2024}. The high melting points of the constituent refractory elements, combined with their elevated strength at high temperatures, make refractory MPEAs attractive for deployment in extreme environments \cite{george_high-entropy_2019}. The kinetics of refractory MPEAs are crucial for identifying promising alloy chemistries suitable for the harsh environments encountered in aerospace or nuclear applications \cite{snead_chapter_2019, wang_designing_2021, tong_influence_2020, xiong_refractory_2023}. For instance, the commercial viability of refractory MPEAs relies on attaining sufficient high-temperature creep resistance, which is significantly influenced by equilibrium vacancy concentrations. Additionally, conventional manufacturing processes face significant challenges in the homogenization of refractory alloys due to slow diffusion even at elevated temperatures.

In this study, we investigate equilibrium vacancy concentrations in a nine-component refractory MPEA comprising elements from groups 4, 5, and 6 of the periodic table. We use first-principles calculations to parameterize an embedded cluster expansion (eCE) model \cite{muller_constructing_2025}. The eCE model, trained on a relatively small training dataset,  accurately reproduces the ordering energetics of all nine elements and vacancies on the body-centered cubic crystal structure. Next, we extend the coarse-graining method of \cite{belak_effect_2015} to rigorously compute finite-temperature equilibrium vacancy concentrations within the canonical ensemble. We employ this scheme to compute equilibrium vacancy concentrations at several equimolar compositions within the nonary alloy. Our rigorous finite-temperature simulations reveal that adding group 4 elements to the multicomponent alloy significantly enhances vacancy concentrations by a factor of 10-100 compared to alloys without group 4 elements. Additionally, our calculations reveal a strong dependence of equilibrium vacancy concentrations on short-range order and bonding energies. This dependence suggests the possibility of tuning alloy compositions to tailor the kinetic properties and microstructural stability of next-generation high-entropy materials.

\section{Statistical thermodynamics of vacancies in multicomponent alloys}
\label{sec:stat_thermo}
The equilibrium vacancy concentration in alloys with a low vacancy content can be rigorously computed with the method outlined by Belak and Van der Ven \cite{belak_effect_2015}. Their formalism calculates the vacancy concentration by averaging the vacancy partition function over orderings of the multicomponent alloy drawn from semi-grand canonical Monte Carlo simulations. Semi-grand canonical simulations of multicomponent alloys can be challenging when the chemical potentials stabilizing compositions of interest are not known a priori. Fixed-composition simulations provide a more practical alternative, particularly in high-dimensional composition spaces containing three or more elements. We begin by deriving the thermodynamic formalism for computing equilibrium vacancy concentrations in multicomponent alloys within the canonical ensemble. The formalism closely follows Belak and Van der Ven's original work \cite{belak_effect_2015}, with some modifications necessitated by the different thermodynamic boundary conditions.

The equilibrium vacancy concentration in a crystalline material with multiple alloying elements is determined by minimizing the total Gibbs free energy with respect to vacancy composition:
\begin{equation}
  \label{eq:equil_vacancy_conventional}
  \mu_{Va}(N_{i},N_{Va}^{*},T,p) = \left(\frac{\partial G}{\partial N_{Va}}\right)_{N_{i},T,p}\bigg|_{N_{Va}=N_{Va}^{*}} = 0
\end{equation}
where $p$ is the pressure, $T$ is the temperature,  $N_{i}$ is the number of atoms of element $i$, $\mu_{Va}$ is the vacancy chemical potential, $G$ is the Gibbs free energy and $N_{Va}^{*}$ is the equilibrium vacancy composition. \Cref{eq:equil_vacancy_conventional} represents the equilibrium condition where the vacancy chemical potential in the bulk crystal equals that in the surrounding environment, where vacancies form readily at defects such as grain boundaries and dislocations. The minimization in \cref{eq:equil_vacancy_conventional} does not conserve the total number of sites in the crystalline region. As simulations with a varying number of crystal sites can be challenging, it is more convenient to compute equilibrium vacancy concentrations while keeping the total number of sites fixed. The differential form of the Gibbs free energy under a fixed total number of sites is given by:
\begin{equation}
  \label{eq:gibbs_fixed_total_sites}
  dG = Vdp - SdT +\sum_{i=2}^{c}\tilde{\mu}_{i} dN_{i} + \mu_{1}dM + \tilde{\mu}_{Va}dN_{Va}
\end{equation}
where $M = \sum_{i=1}^{c}N_{i} + N_{Va}$ is the total number of sites, and $\tilde{\mu}_{i} = \mu_{i} - \mu_{1}$ are the exchange chemical potentials for each chemical specie relative to specie $1$. Under these conditions, the equilibrium number of vacancies is given by:
\begin{align}
  \label{eq:equil_vacancy_fixed_total_sites}
  \tilde{\mu}_{Va}(N_{i\neq 1},M,N_{Va}^{\dagger},T,p) &= \left(\frac{\partial G}{\partial N_{Va}}\right)_{N_{i\neq 1},M,T,p}\bigg|_{N_{Va}=N_{Va}^{\dagger}} \notag \\
                                            &= -\mu_{1}
\end{align}
In this case, the Gibbs free energy is not minimized as in \cref{eq:equil_vacancy_conventional}, but instead reaches a value where its derivative with respect to vacancy number equals the negative chemical potential of species 1. The vacancy concentrations calculated through both \cref{eq:equil_vacancy_conventional} and \cref{eq:equil_vacancy_fixed_total_sites} should be nearly identical in the thermodynamic limit when vacancy concentrations are small. 

The numerical evaluation of \cref{eq:equil_vacancy_fixed_total_sites} is challenging as vacancy concentrations in technologically important alloys are extremely low. The equilibrium vacancy concentration is easier to measure in a thermodynamic ensemble where the vacancy exchange chemical potential ($\tilde{\mu}_{Va}$) is held fixed instead of the number of vacancies. Under such boundary conditions, the differential form of the relevant free energy is:
\begin{equation}
  \label{eq:char_potential_differential}
  d\Phi =  Vdp - SdT +\sum_{i=2}^{c}\tilde{\mu}_{i} dN_{i} + \mu_{1}dM - N_{Va} d\tilde{\mu}_{Va}
\end{equation}
The number of vacancies at equilibrium, $N_{Va}^{\dagger}$ is then given by:
\begin{equation}
  \label{eq:equil_vacancy_char_potential}
  N_{Va}^{\dagger} = \left(\frac{\partial \Phi}{\partial \tilde{\mu}_{Va}}\right)_{N_{i\neq 1},M,T,p}\bigg|_{\tilde{\mu}_{Va}=-\mu_{1}}
\end{equation}
\Cref{eq:equil_vacancy_char_potential} enables the estimation of equilibrium vacancy concentration ($x_{Va} = \frac{N_{Va}^{\dagger}}{M}$) at fixed alloy composition, provided $\mu_{1}$ is known.

The number of vacancies at equilibrium (\cref{eq:equil_vacancy_char_potential}) can be directly computed with statistical mechanics techniques. The partition function associated with the free energy $\Phi$ is:
\begin{equation}
  \label{eq:partition_function}
  Z = \sum_{\vec{\sigma}}\sum_{N_{Va}=0}^{N_{Va}=M-\sum_{i}N_{i\neq 1}}\sum_{\vec{\nu}}\exp(-\beta \Omega(\vec{\nu}))
\end{equation}
where $\vec{\sigma}$ represents all possible arrangements of $N_{i\neq 1}$ atoms of element $i$, and $N_{1} = M-\sum_{i=2}^{c}N_{i}$ atoms of element 1 over the $M$ crystal sites. $N_{Va}$ denotes the number of vacancies, and $\vec{\nu}$ represents all possible vacancy arrangements obtained by replacing $N_{Va}$ atoms of element 1 with vacancies. The vacancy ordering $\vec{\nu}$ depends on the overall state of ordering, $\vec{\sigma}$. The energy, $\Omega(\vec{\nu})$, in an alloy with element compositions of $x_{i}=\frac{N_{i}}{M}$ and ordering $\vec{\nu}$ is:
\begin{equation}
  \label{eq:vacancy_formation_energy}
  \Omega(\vec{\nu}) = E(\vec{\nu}) - \tilde{\mu}_{Va}N_{Va}
\end{equation}
\Cref{eq:partition_function} first counts over all possible decorations of $M$ sites with a fixed number of $N_{i\neq 1}$ atoms. Since the number of vacancies can fluctuate in this ensemble, the partition function of \cref{eq:partition_function} sums over all vacancy numbers ranging from zero to $N_{1}$, followed by all possible arrangements of the $N_{Va}$ vacancies over the $N_{1}$ sites occupied by element 1.

Metallic alloys typically have large vacancy formation energies leading to low vacancy concentrations. For materials with low vacancy concentrations and when the number of sites ($M$) is not excessively large, we can simplify \cref{eq:partition_function} by neglecting configurations with two or more vacancies:
\begin{equation}
  \label{eq:partition_function_simplified}
  Z \approx \sum_{\vec{\sigma}} \exp(-\beta \Omega(\vec{\sigma}))\left[1+\sum_{\vec{\nu}}\exp(-\beta \Delta\Omega(\vec{\nu}))\right]
\end{equation}
where $\Delta \Omega(\vec{\nu}) = E(\vec{\nu}) - E(\vec{\sigma}) -\tilde{\mu}_{Va}$, which is the vacancy formation energy of ordering $\vec{\nu}$. $E(\vec{\sigma})$ is the energy of an ordering without vacancies and $E(\vec{\nu})$ the energy of an ordering where one atom of specie 1 is replaced by a vacancy. Following the approach of Belak and Van der Ven \cite{belak_effect_2015}, we define $\tilde{Z}_{\textrm{alloy}} = \sum_{\vec{\sigma}} \exp(-\beta \Omega(\vec{\sigma}))$ as the canonical partition function of the vacancy-free alloy, and:
\begin{align}
  \label{eq:xi_definition}
  \xi &= \sum_{\vec{\sigma}} \frac{\exp(-\beta \Omega(\vec{\sigma}))} {\tilde{Z}_{\textrm{alloy}}} \sum_{\vec{\nu}}\exp(-\beta \Delta\Omega(\vec{\nu}))\nonumber \\
      &= \langle \sum_{\vec{\nu}}\exp(-\beta \Delta\Omega(\vec{\nu})) \rangle_{\textrm{alloy}}
\end{align}
where $\xi$ represents the ensemble-averaged vacancy partition function, computed over microstates of the vacancy-free alloy. The partition function of \cref{eq:partition_function_simplified} is then given by:
\begin{equation}
  \label{eq:parition_function_xi}
  Z \approx \tilde{Z}_{\textrm{alloy}} (1+\xi)
\end{equation}

The number of vacancies at equilibrium, $N_{Va}^{\dagger}$, can be estimated as an ensemble average over all microstates accessible to the system. Assuming the number of vacancies is limited to at most one and that the crystalline region is sufficiently large to be in the thermodynamic limit, \cref{eq:parition_function_xi,eq:partition_function_simplified,eq:xi_definition} can be combined to obtain an expression for the equilibrium number of vacancies:
\begin{equation}
  \label{eq:approx_vacancy_concentration}
   N_{Va}^{\dagger} \approx \left(\frac{\xi}{1+\xi}\right)
\end{equation}
The vacancy concentration is then obtained as $x_{Va} = \frac{N_{Va}}{M}$. \Cref{eq:approx_vacancy_concentration} is identical to the expression derived by Belak and Van der Ven \cite{belak_effect_2015}, with a slight difference in the definitions for $\xi$ arising from the difference in ensembles.

It is important to note that the above formalism applies only to dilute vacancy concentrations, i.e. $\xi\ll1,  N_{Va}^{\dagger}\ll1$. For materials with large vacancy concentrations, $N_{Va}^{\dagger}$ can approach or exceed unity. The calculated vacancy concentrations are inaccurate in such cases, as vacancy numbers larger than 1 have been neglected in the derivation of \cref{eq:approx_vacancy_concentration}.

As shown in \cref{sec:appendix_SRO_vacancy}, the number of atoms of element $i$ around a vacancy is given by:
\begin{equation}
  \label{eq:local_concentration_approx}
  \langle N_{\alpha}^{i} \rangle \approx \frac{\theta_{\alpha}^{i}}{1+\xi}
\end{equation}
where $\theta_{\alpha}^{i} = \langle \sum_{\vec{\nu}}N_{\alpha}^{i}(\vec{\nu})\exp(-\beta \Delta\Omega(\vec{\nu}))\rangle_{\textrm{alloy}}$. The local concentration of element $i$ within a specific coordination shell $\alpha$ can then be computed as $x_{\alpha}^{i} = \frac{\theta_{\alpha}^{i}/\xi}{N_{\alpha}}$ where $N_{\alpha}$ is the number of sites in the coordination shell.

\Cref{eq:approx_vacancy_concentration,eq:local_concentration_approx} can be evaluated with canonical Monte Carlo simulations performed on an alloy containing $N_{i\neq 1}$ atoms of element $i$ and $M-\sum_{i=2}^{c}N_{i}$ atoms of specie 1 at a temperature $T$. Finite-temperature orderings are drawn from the Monte Carlo trajectory, and for each configuration, the sum over vacancy formation energies is computed by replacing every site containing element 1 with a vacancy. The ensemble average value of the vacancy partition functions sampled across the entire Monte Carlo trajectory is then used to compute $\xi$ through \cref{eq:xi_definition}. Similarly, the values of $\theta_{\alpha}^{i}$, representing the local concentration of element $i$ weighted by the vacancy formation energy, are computed along the same Monte Carlo trajectory. \Cref{eq:approx_vacancy_concentration,eq:local_concentration_approx} can subsequently be used to determine both the equilibrium vacancy concentration and the local concentration of various elements surrounding the vacancy.

Methods such as the ``density of vacancy formation states'' (VF-DOS) proposed by Zhang \emph{et al.} \cite{zhang_ab_2022} to compute equilibrium vacancy concentrations can be derived from \cref{eq:approx_vacancy_concentration} by imposing additional approximations. When $\xi$ is sufficiently small ($\xi\ll1$), the denominator of \cref{eq:approx_vacancy_concentration} can be neglected:
\begin{equation}
  \label{eq:approx_1_zhang}
  x_{Va}=\frac{\frac{1}{M}\xi}{1+\xi}\approx \frac{1}{M}\xi = x_{1} \langle \exp(-\beta\Delta\Omega(\vec{\nu})) \rangle_{\textrm{alloy},\textrm{crystal}}
\end{equation}
where $\langle \exp(-\beta\Delta\Omega(\vec{\nu})) \rangle_{\textrm{alloy},\textrm{crystal}} = \frac{\xi}{N_{1}}$ is the average value of $\exp(-\beta\Delta\Omega(\vec{\nu}))$ over all possible orderings and vacancy exchanges. Additionally, defining $\Delta\epsilon_i(\nu) = E(\vec{\nu})-E(\vec{\sigma})+\mu_i^{ex}$, where the excess chemical potential of specie $i$ is $\mu_{i}^{ex} = \mu_i-\mu_i^{ideal} =\mu_i-k_BT\ln{x_i}$, the equilibrium vacancy concentration can be written as:

\begin{equation}
  \label{eq:approx_zhang}
  x_{Va} = \langle \exp\left({-\beta\Delta\epsilon(\vec{\nu})}\right) \rangle_{\textrm{alloy},\textrm{crystal}}
\end{equation}
Replacing the chemical potential with the excess chemical potential removes the factor of $x_{1}$ and yields an expression mirroring the VF-DOS formulation of \cite{zhang_ab_2022}. \Cref{eq:approx_zhang} is derived by replacing only atoms of species 1 with vacancies. Alternatively, the ensemble average in \cref{eq:approx_zhang} can also be computed by sequentially replacing each element with a vacancy, provided that appropriate excess chemical potentials are used in the computation of $\Delta\epsilon$. In studies such as \cite{zhang_ab_2022, li_first_2019, mizuno_defect_2019, guan_chemical_2020}, \cref{eq:approx_zhang} is typically evaluated using only a single configuration, which is usually drawn from an approximation of a disordered solid solution such as a special quasi-random structure (SQS).

\section[methods]{Methods}
\label{sec:methods}

Estimating vacancy concentrations with \cref{eq:approx_vacancy_concentration} requires an accurate model for the energy of different atomic arrangements of the $c$ chemical species and vacancies on a parent crystal structure, as well as the chemical potential $\mu_{1}$ of species 1 to compute the vacancy formation energy in \cref{eq:vacancy_formation_energy}. \emph{Embedded} cluster expansions (eCE) \cite{muller_constructing_2025}, parameterized using first-principles calculations, serve as surrogate models to estimate the ordering energetics of multicomponent alloys with vacancies. To estimate the chemical potentials of disordered solid solutions, we use a combination of free energy integration and the Widom particle exchange technique \cite{widom_topics_1963}. We demonstrate the computation of vacancy concentrations and chemical trends in a prototypical 9-component refractory MPEA system composed of elements from groups 4, 5, and 6 of the periodic table. At elevated temperatures, alloys formed from these 9 elements predominantly adopt a disordered solid solution on the bcc crystal structure \cite{natarajan_crystallography_2020}.

\subsection{Embedded Cluster Expansions (eCE)}

Embedded cluster expansions (eCE) \cite{muller_constructing_2025} provide computationally efficient atomistic models to describe the ordering energetics of  multicomponent alloys containing several alloying elements. eCE models partition the energy into contributions arising from each site in the crystal:
\begin{equation}
  \label{eq:eCE_energy}
  E(\vec{\sigma}) = \sum_{i\in M} E_{i}(\vec{\sigma})
\end{equation}
where $\vec{\sigma}$ represents the occupants of the $M$ sites in the crystal, $E$ is the total energy of the crystal, and $E_{i}$ is the energy contribution from site $i$. The site energy depends on the chemical species that occupy sites in the neighborhood of site $i$.

The site occupancy is mathematically described using $c$ site basis functions, $\vec{\phi(\sigma_{i})}$. eCE models project the site basis functions into a lower dimensional space through a linear transformation:
\begin{equation}
  \label{eq:eCE_site_basis_functions}
  \vec{\tilde{\phi}}(\sigma_i) = T\vec{\phi}(\sigma_i)
\end{equation}
The matrix $T$ is a learnable embedding matrix, and $\vec{\tilde{\phi}}$ are the embedded site basis functions. Local descriptors of ordering are constructed by taking tensor products of site basis functions across all sites within a cutoff distance of the central site. The tensor products are subsequently symmetrized with the symmetry group of the disordered parent crystal to create symmetry-invariant local descriptors of ordering. Neural networks use these symmetrized descriptors as inputs to predict the site energies and the total energy of a crystal through \cref{eq:eCE_energy}. The eCE formalism enables the fast and accurate prediction of formation energies across diverse atomic configurations. eCE models have been shown to leverage chemical similarities between alloying elements to reduce the complexity of atomistic models and require fewer electronic structure calculations to parameterize than traditional cluster expansion models \cite{muller_constructing_2025}.

The formation energies of configurations in the training dataset were computed with density functional theory (DFT) as implemented in the Vienna Ab-initio Simulation Package (VASP) \cite{kresse_efficiency_1996, kresse_efficient_1996}. The generalized gradient approximation (GGA)-type Perdew-Burke-Ernzerhof (PBE) functional was used to approximate the unknown exchange-correlation contribution in DFT \cite{perdew_generalized_1996}, with the projected augmented wave (PAW) potentials for the description of core electrons. \cite{kresse_ultrasoft_1999} Valence electrons were expanded in a plane-wave basis set with an energy cutoff of 550~eV. A $\Gamma$-centered Monkhorst-Pack  $k$-point mesh with a grid density of 55~Å ensured accurate Brillouin zone sampling \cite{monkhorst_special_1976}. The total energy of each structure was converged to within 10$^{-4}$~eV/cell. The projected augmented wave (PAW) potentials used are listed in Table \ref{SI:table:potcar} of the Supplemental Material.

eCE models are trained on symmetrically distinct arrangements of vacancies and elements from groups 4, 5, and 6 of the periodic table. The dataset contains various orderings of the ten different species on the bcc crystal structure. Orderings of the nine refractory elements in small cells (containing up to 27 atoms) are included to accurately reproduce the thermodynamics of the refractory alloy without vacancies. The energies of 8179 configurations were calculated to ensure sufficient coverage of the nonary composition space. 

The dataset also includes configurations with dilute vacancy concentrations to capture vacancy energetics. Vacancy energies are computed in large supercells to minimize vacancy-vacancy interactions. Single vacancies inserted in special quasi-random structures \cite{zunger_special_1990} approximate the effect of disorder on vacancy thermodynamics. The dataset contains SQS-based orderings enumerated in a $4\times 4\times 4$ supercell of the primitive bcc cell. These configurations contain between 2-6 refractory elements. The dataset also contains vacancies included in SQS structures for the Senkov \cite{senkov_mechanical_2011} alloy composition (HfNbTaTiZr) and the equimolar 9-component refractory alloy. Vacancy configurations comprise 1,240 data points spanning the entire 9-component space.

Structures with lattice and basis deformation costs larger than 0.015 as computed with the mapping algorithm described in \cite{thomas_comparing_2021} are excluded from the dataset. We randomly split the final dataset into train, test, and validation sets in an 8:1:1 ratio. Orderings of the pure elements, and dilute defects of solute and vacancies within the pure elements are always included in the training dataset. We maintained a consistent proportion of vacancy-containing structures across all datasets.

A 4-eCE model was parameterized on the training dataset by minimizing the mean squared error with the Adam algorithm. Features for the eCE model were derived based on cluster functions composed of pairs and triplets within distances of 7.8 and 4~Å of the central site. The final clusters included in the model consist of 1 empty cluster, 1 point cluster, 7 pair clusters, and 1 triplet cluster. Overfitting was reduced through the implementation of early stopping based on the validation loss and L2 regularization of the model parameters. The energy network for the eCE model comprised of a 5-layer neural network ($128\times128\times32\times8\times1$) in which the ReLU activation function was used in all layers except the readout layer.

The eCE model was trained to reproduce the formation energy per primitive cell, calculated as:
\begin{equation}
    \Delta e_f= \frac{E(C^{1}_{n_{1}}C^{2}_{n_{2}}\cdots C^{c}_{n_{c}})-\sum_{i=1}^{c}n_{i} E(C^{i})}{\sum_{i=1}^{c}n_{i}}
\end{equation}
where $E(C^{1}_{n_{1}}C^{2}_{n_{2}}\cdots C^{c}_{n_{c}})$ is the energy of the alloy configuration and $E(C^{i})$ is the energy of the reference states as calculated by DFT. $C^i$ represents species $i$ (including vacancies) and $n_i$ represents the number of species $C^i$. The reference structures used were the BCC structures of all pure elements.

\subsection{Canonical Monte Carlo simulations}
\label{sec:Monte Carlo-simul}

Ensemble averages for estimating vacancy thermodynamics are computed from canonical Monte Carlo simulations performed with the Metropolis algorithm \cite{metropolis_equation_1953} in a $10\times10\times10$ supercell (2000 atoms) of the conventional bcc structure. The simulation is initialized at a temperature of 10,000~K and cooled to a final temperature of 1000~K. Monte Carlo simulations are allowed to run until the ensemble averages of formation energy are converged to a precision of 1 meV/site with a confidence level of 95\%. The canonical Monte Carlo simulations are performed at alloy compositions that do not contain any vacancies. Uncorrelated samples are then extracted \cite{shirts_statistically_2008, chodera_use_2007} to compute the Gibbs free energies, exchange chemical potentials, and vacancy concentrations at elevated temperatures.

The chemical potential in \cref{eq:vacancy_formation_energy} is computed with the Widom substitution method and free energy integration. Exchange chemical potentials of specie $i$ with respect to a reference specie (denoted 1 here) are given by:
\begin{equation}
  \label{eq:widom}
    \tilde{\mu}_{i}=-k_BT\ln{\frac{N_1}{N_i+1}}
    -k_BT\ln{\langle e^{-\beta \Delta E}
    \rangle_{N_{1},N_{2},\cdots,N_{c}}}
\end{equation}
where $\Delta E$ is the energy difference due to the replacement of one atom of the reference element, with specie $i$. A detailed derivation of \cref{eq:widom} is shown in \cref{sec:append-widom-subst}. Tests shown in \cref{sec:results_finite_temp} indicate that the vacancy concentrations are insensitive to the choice of the reference specie.

The Gibbs free energy of the disordered solid solution at temperature $T$ and composition $x_{i} = \frac{N_{i}}{M}$ is calculated using free energy integration \cite{walle_self-driven_2002, puchala_casm_2023}:
\begin{equation} \label{eqn:free_energy_integration}
    \beta g(T,x_{i})= \beta_0 g(T_0,x_{i})+
    \int_{\beta_0}^\beta \langle e\rangle d\beta
\end{equation}
where $\beta=1/k_BT$, $\langle e\rangle$ is the ensemble average of the formation energy per atom at temperature $T$, $g(T,x_{i}) = \frac{G(T,N_{i})}{M}$ is the Gibbs free energy per atom, and the integral is computed from a high temperature $T_0$ to $T$. The free energy of the disordered solid solution at a high temperature $T_{0}$ is approximated using the disordered enthalpy from Monte Carlo simulations and ideal solution entropy.

The chemical potential of specie 1 is computed from the free energy in \cref{eqn:free_energy_integration} and the exchange chemical potentials of \cref{eq:widom} as the intercept of the tangent plane to $g(x_{i},T)$ with the $x_{1}=1$ axis:
\begin{equation}
  \label{eq:chem_pot_1}
  \mu_{1}=g+\sum_{i=2}^c \tilde{\mu}_i(-x_i)
\end{equation}

\section{Results}

Multicomponent alloys of groups 4, 5, and 6 of the periodic table are attractive candidates for high-temperature structural applications and deployment in extreme environments \cite{miracle_strength_2024}. These materials primarily adopt the body-centered cubic (bcc) crystal structure, though depending on the elements and their precise ratios, phases with the hexagonal close-packed (hcp), or more complex precipitates with the Laves crystal structure may form \cite{natarajan_crystallography_2020}. Processing and synthesizing multicomponent refractory alloys are especially challenging due to sluggish kinetics. The origin of this sluggish kinetics could arise due to several reasons including low vacancy concentrations, or large atomic diffusion barriers. We analyze the vacancy concentrations in the nine-component alloy comprised of elements from groups 4, 5, and 6 of the periodic table.

\subsection{Parameterization of eCE model}
\begin{figure}[h]
    \centering
    \includegraphics{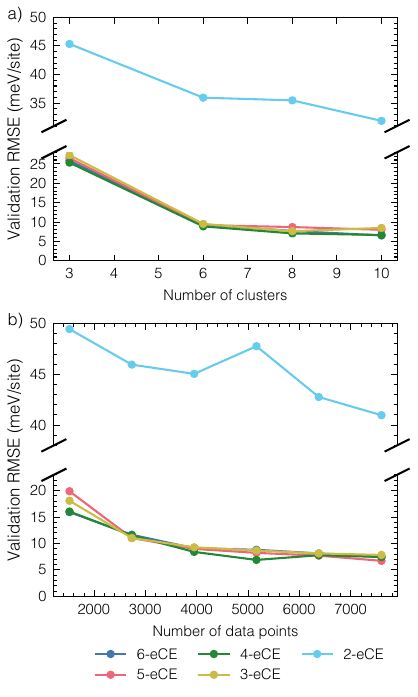}
    \caption{Learning curves of $k$-eCE ($k$=2-6) models showing the (a) variation in the validation error with number of clusters and (b) variation in validation error with number of data points. In (b), the data points were sequentially added based on their maximum Mahalanobis distance from the existing dataset. For each addition of data, the ratio of configurations containing no vacancies to those containing vacancies were kept constant.}
    \label{fig:learning_curves}
\end{figure}

Formation energies of 9419 configurations were used to parameterize the eCE models of \cref{fig:learning_curves}. The validation errors in \cref{fig:learning_curves}a are based on models trained on 7595 configurations and validated against a separate set of 912 configurations. Validation errors are obtained from eCE models with embedding dimensions ranging from 2 to 6 in \cref{fig:learning_curves}a. 2-eCE models have very large validation errors exceeding $\approx 30$ meV/site, while models with embedding dimensions between 3-6 have very similar errors. Pair clusters up to the 4th nearest neighbor accurately reproduce the formation energies to within $\approx 10$~meV/site. Adding the nearest neighbor triplet cluster along with a few additional pairs for a total of 10 clusters lowers the validation error to 6-7~meV/site. \Cref{fig:learning_curves}b displays the variation in the validation error of the eCE models on the number of training data points. A dataset size of $\approx 4000$ training data points lowers the validation errors of the eCE models to $\approx 9$~meV/site. The low prediction error of eCE models with relatively small cluster sizes and training datasets is likely due to the versatility of eCE models to learn chemical similarities and leverage non-linear energy models to accurately capture ordering energetics.

Obtaining accurate fits of eCE models requires careful initialization of the embedding matrix, $T$, in \cref{eq:eCE_site_basis_functions}. For our eCE fits, we initialize the embedding matrix following the scheme proposed in \cite{muller_constructing_2025}, which employs the chemical properties of elements to compute initial values based on singular value decomposition. All chemical properties of the vacancy, which is treated as a chemical species in our eCE model, are set to zero. The learned embedding matrix exhibits clustering of chemically similar elements within single groups (i.e., Zr and Hf, Mo and W, Nb and Ta), while separating the 3d refractory metals, consistent with previous findings \cite{muller_constructing_2025}. Notably, vacancies appear well-separated from the refractory elements in the embedding space, indicating their chemical dissimilarity and suggesting that vacancy interactions are distinct from all other chemical interactions in the system.

Based on the analysis of \cref{fig:learning_curves} and model complexity versus performance (\cref{SI:fig:tradeoff}), we selected a 4-eCE model as it provides an optimal balance between computational efficiency and predictive accuracy. This model achieves an overall RMSE of 7.72 meV/site on an unbiased test. Configurations without vacancies show an RMSE of 7.96 meV/site, while those containing vacancies exhibit a lower RMSE of 5.44 meV/site. A comparison between the formation energies predicted by the 4-eCE model and DFT is shown in \cref{SI:fig:parity_plot} in the Supplemental Material. 

\begin{figure}
    \includegraphics{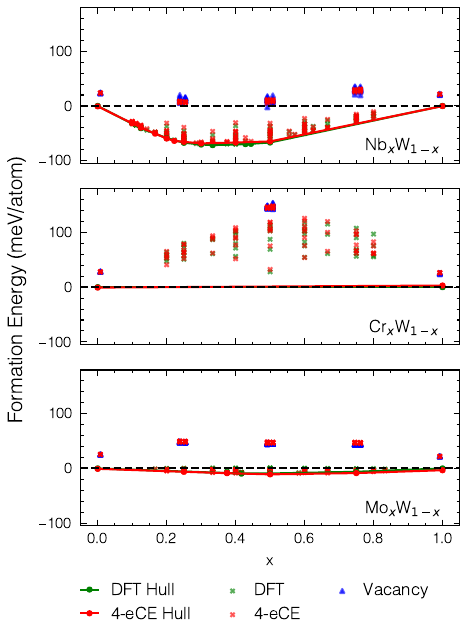}
    \caption{Convex hulls of 3 binary alloys (CrW, NbW and MoW) predicted by DFT and the 4-eCE model. Formation energies calculated using DFT and predicted by the CE model are shown in green and red respectively. The DFT formation energies of the configurations containing vacancies are displayed in blue triangles.}
    \label{fig:convex_hulls}
  \end{figure}

\begin{figure}
    \includegraphics{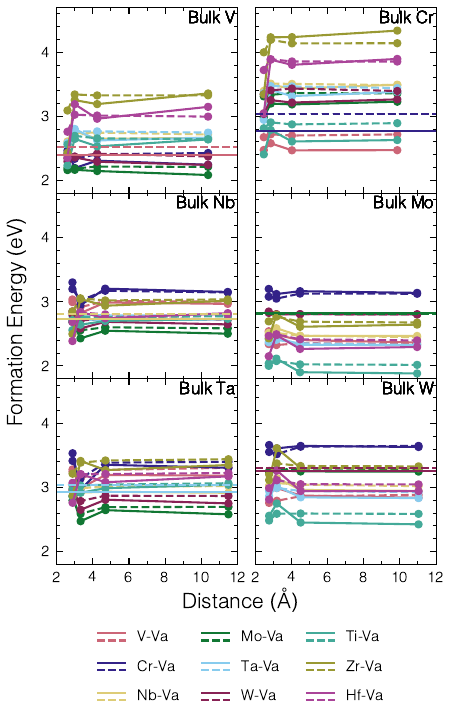}
    \caption{Formation energies of a solute atom and a vacancy dissolved in a bulk solvent ($4\times4\times4$ BCC supercell), as a function of the distance between the solute and the vacancy. The solvent element in the bulk system are indicated within each figure. The vacancy formation energies in the solvent are displayed as horizontal lines. Solid lines represent energies calculated using DFT while the dashed lines represent energies predicted by the 4-eCE model. Each solute-vacancy combination is denoted by a different color.}
    \label{fig:solute_vacancy}
\end{figure}

\begin{figure}
    \centering
    \includegraphics{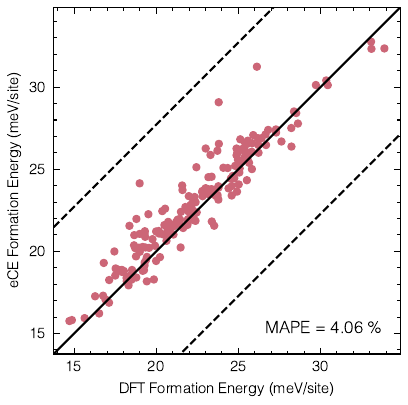}
    \caption{Parity plot comparing the 4-eCE predicted and DFT-calculated energies of solute-vacancy configurations across all solvents. The model error (RMSE) is indicated by the dashed lines. The mean absolute percentage error (MAPE) is also shown in the bottom right of the plot.}
    \label{fig:solute_vacancy_parity}
  \end{figure}

  For further validation, we performed Canonical Monte Carlo simulations on a $4\times 4\times 4$ supercell of bcc with three compositions: HfNbTaTiZr, TiZrHfVNbTaCrMoW, and ZrNbMo, each containing a single vacancy. DFT calculations on the ten configurations with the lowest formation energy from each composition revealed an RMSE of 7.20 meV/site between DFT and the eCE model. As this error is within the numerical accuracy of our 4-eCE model, we consider the atomistic model to be sufficiently reliable.

  The 4-eCE model's accuracy in predicting binary alloy ordering energetics is evaluated by constructing convex hulls for all thirty-six refractory binary systems. \Cref{fig:convex_hulls} shows the convex hulls for Cr-W, Nb-W, and Mo-W systems. Comparisons for the remaining systems are presented in \cref{SI:fig:convex_hull_group4,SI:fig:convex_hull_group5-6}. The 4-eCE model accurately reproduces the shape of all binary convex hulls and predicts the formation energies of vacancy-containing configurations. 

  A closer comparison of solute-vacancy binding energies, presented in \cref{fig:solute_vacancy,fig:solute_vacancy_parity}, shows general agreement between the 4-eCE model and DFT. Each panel within \cref{fig:solute_vacancy} compares the extensive formation energies of solute-vacancy pairs separated by varying distances while embedded in different solvents. For example, DFT calculations suggest Ti-Va pairs in molybdenum repel each other at short distances and prefer larger separations. This can be seen in the DFT energies of \cref{fig:solute_vacancy} (indicated by the solid lines), where the energy of Ti-Va pairs separated by $\approx$ 11~Å is lower than that of a nearest-neighbor or next nearest-neighbor Ti-Va pair. The 4-eCE model, represented by the dashed line in \cref{fig:solute_vacancy}, predicts the same trend. In fact, the 4-eCE model also accurately reproduces the slightly stronger repulsion observed for Ti-Va pairs at the second nearest-neighbor distance compared to the first nearest-neighbor distance. Similar qualitative effects related to solute-vacancy separation are reproduced by the model across all solvents. The comparison between the DFT-calculated and the 4-eCE predicted formation energies are shown in \cref{fig:solute_vacancy_parity}. A root mean square error (RMSE) of 1.2 meV/site, or a mean absolute percentage error (MAPE) of 4.06\%, is achieved by these configurations.” Slight discrepancies between the model and DFT do arise. These can be largely attributed to small errors in the model's prediction of either the solute dissolution energy or the vacancy formation energy within the solvent.

\subsection{Finite-temperature simulations}
\label{sec:results_finite_temp}

\begin{figure}
\includegraphics{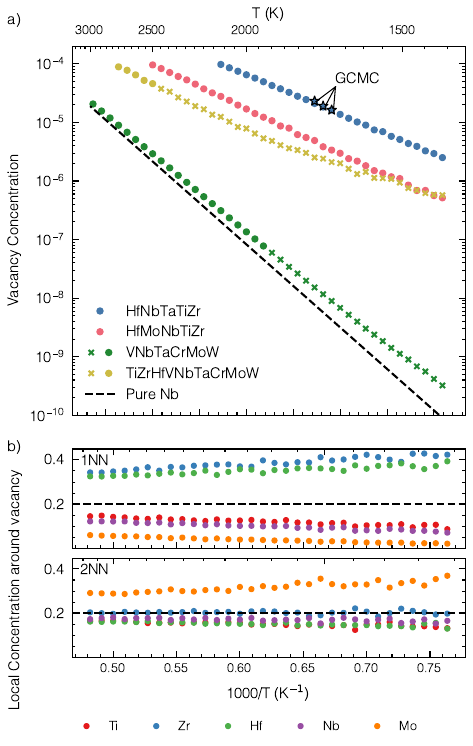}
    \caption{a) Plot of $x_{Va}$ versus $1000/T$ (bottom axis) and $T$ (top axis) for 4 different alloys. The vacancy concentrations obtained from GCMC for the HfNbTaTiZr alloy are superimposed on the plot as blue stars. Crosses indicate that the alloy could undergo phase separation into multiple phases.  b) Average local concentration around each vacancy in the HfMoNbTiZr alloy, for the \nth{1} and \nth{2} nearest neighbors (indicated by 1NN, 2NN). The black dashed line indicates the average composition of each species in the alloy.}
    \label{fig:vacancy_conc}
  \end{figure}

  Finite-temperature vacancy concentrations in the nonary refractory alloy were computed using Monte Carlo simulations based on our 4-eCE model. \Cref{fig:vacancy_conc}a shows the equilibrium vacancy concentration as a function of inverse temperature for four equiatomic multicomponent alloys: HfMoNbTiZr and HfNbTaTiZr (two widely studied refractory alloys in the literature \cite{miracle_critical_2017,miracle_strength_2024}), VNbTaCrMoW (containing elements from groups 5 and 6 of the periodic table), and the equiatomic nonary alloy TiZrHfVNbTaCrMoW.

  \Cref{fig:vacancy_conc}a illustrates the effects of alloy composition and temperature on equilibrium vacancy concentrations. The equiatomic 6-component VNbTaCrMoW alloy exhibits the lowest equilibrium vacancy concentration among the four alloys. The large vacancy formation energies of the group 5 and 6 elements likely contributes significantly to this alloy's depressed vacancy concentration. In contrast, the other alloys shown in \cref{fig:vacancy_conc}a with substantial concentrations of group 4 elements are predicted to have vacancy concentrations between $10^{-6}-10^{-4}$. The vacancy concentrations of HfMoNbTiZr, HfNbTaTiZr, and TiZrHfVNbTaCrMoW are two to three orders of magnitude higher than those of the alloy containing only group 5 and 6 elements. In fact, substituting Zr for any element in the equiatomic VNbTaCrMoW base alloy results in a higher vacancy concentration (\cref{SI:fig:vac_conc_substitution}). The nearly linear relationship between vacancy concentration and inverse temperature in \cref{fig:vacancy_conc}a indicates Arrhenius-like behavior in these refractory alloys, with deviations arising from phase instability at lower temperatures.

  The calculations in \cref{fig:vacancy_conc}a reveal several chemical trends. The equiatomic senary alloy comprised of elements from group 5 and 6 has a vacancy formation energy of around 2.76~eV. This vacancy formation energy is close to that of pure niobium ($\approx 2.81$~eV in \cref{fig:solute_vacancy}), suggesting that niobium can serve as an effective surrogate or ``average'' element to represent the other elements of groups 5 and 6. 

  Interestingly, our calculations in \cref{fig:vacancy_conc}a indicate that substituting tantalum for molybdenum in the 6-component HfMoNbTiZr can increase vacancy concentration by almost an order of magnitude. This is unexpected as the vacancy formation energy of pure tantalum is higher than that of pure molybdenum. The equiatomic nonary alloy, containing higher concentrations of group 5 and 6 elements, has a lower vacancy concentration than either of the quinary alloys of \cref{fig:vacancy_conc}a. This suggests that group 4 elements are important additions to refractory alloys for enhancing vacancy concentrations.

  The average local concentrations of each element around vacancies in the HfMoNbTiZr alloy are shown in \cref{fig:vacancy_conc}b. Our simulations predict an enhanced concentration of Zr and Hf in the first nearest-neighbor shell surrounding the vacancy across the entire temperature range. The second nearest-neighbor coordination environment is predicted to be richer in Mo. These local elemental segregation trends align with the dilute solute-vacancy binding preferences presented in \cref{fig:solute_vacancy} and \cref{SI:fig:solute_vacancy_split}. Both Hf and Zr are predicted to strongly bind vacancies in the first-neighbor shell for all the solvents considered in \cref{fig:solute_vacancy}. Similarly, Mo shows a preference for the second-neighbor shell when Nb is the solvent. Thus, the energetic preferences for solute-vacancy binding identified in dilute alloys (\cref{fig:solute_vacancy}) are correlated with the elemental segregation observed in \cref{fig:vacancy_conc}b. However, although trends from dilute alloys can offer qualitative guidance, they may be quantitatively unreliable for MPEAs because the complex local environments and competing interactions can lead to different behavior.

  \Cref{fig:vacancy_conc}a validates the formalism introduced in \cref{sec:stat_thermo} by comparing the predictions of \cref{eq:approx_vacancy_concentration} against semi-grand canonical Monte Carlo (GCMC) simulations. In the GCMC simulations, the exchange chemical potentials of all species, including vacancies, are fixed to the values computed via the Widom particle exchange method. The comparison is carried out for the HfNbTaTiZr alloy, which exhibits a sufficiently high vacancy concentration to enable direct estimation of equilibrium vacancy levels in long-duration GCMC simulations. As shown in \cref{fig:vacancy_conc}a, the equilibrium vacancy concentration obtained from GCMC simulations aligns closely with the value predicted by the coarse-graining approach described in \cref{sec:stat_thermo}. The alloy composition measured in GCMC simulations was (to within numerical tolerance) equal to that of the equimolar alloy. Interestingly, while GCMC required approximately $\approx 10^8$ samples to achieve statistically accurate vacancy concentrations, our coarse-graining method reached comparable accuracy with only a few hundred samples.

Our formalism, described in \cref{sec:stat_thermo}, allows for vacancy swaps involving a single species in the alloy. The results of \cref{fig:vacancy_conc}a are independent of the element that is replaced by a vacancy to compute the vacancy concentration in \cref{eq:xi_definition}. \Cref{SI:fig:vac_conc_diff_swap} estimates the equilibrium vacancy concentration by replacing each of the five elements in the HfNbTaTiZr alloy with a vacancy. The results, plotted together in \cref{SI:fig:vac_conc_diff_swap}, are independent of the element choice and demonstrate that all five cases yield essentially identical vacancy concentrations.

Computing vacancy concentrations with the formalism of \cref{sec:stat_thermo} can be challenging when the equilibrium number of vacancies in the simulation cell exceeds 1 or if the system undergoes an order-disorder phase transformation. To mitigate instances where the vacancy concentration is too large to apply the approximations of \cref{sec:stat_thermo}, we do not show any vacancy concentrations in \cref{fig:vacancy_conc} exceeding $2\times10^{-4}$ as this value approaches the limit of our dilute vacancy approximation. Additionally, both the VNbTaCrMoW and TiZrHfVNbTaCrMoW alloys are predicted to undergo a phase transition at the temperatures shown in \cref{fig:vacancy_conc}a. Phase transitions are inferred based on the heat capacities from MC simulations shown in \cref{SI:fig:heat_capacity}. The senary VNbTaCrMoW exhibits a clear phase transition, evidenced by a divergence in heat capacity near 1400~K. For the nonary TiZrHfVNbTaCrMoW alloy, the presence and onset of a phase transition is less distinct. GCMC simulations performed with chemical potentials estimated from canonical simulations indicate significant deviations from equiatomic composition at temperatures below $\sim 2600$~K for the nonary alloy and $\sim 1900$~K for the senary alloy, suggesting an order-disorder phase transition below these temperatures. Vacancy concentrations computed below temperatures where the system has undergone a phase transition represent the vacancy concentration of multi-phase mixtures rather than those of individual phases. These vacancy concentrations should therefore be interpreted with caution. To reflect this uncertainty, we have explicitly marked the temperature range where the alloys may not remain a stable single phase in \Cref{fig:vacancy_conc}a with crosses. These results further suggest the presence of a significant miscibility gap in the 9-component equiatomic alloy, persisting even at elevated temperatures and contradicting the common assumption that configurational entropy alone is sufficient to stabilize a single-phase solid solution in high-entropy alloys.

\section{Discussion}

Vacancy concentrations are important inputs to predict the kinetic properties of concentrated alloys. This study presents a rigorous first-principles statistical mechanics analysis of vacancy concentrations in a nine-component alloy of elements from groups 4, 5, and 6 of the periodic table. The embedded cluster expansion (eCE) \cite{muller_constructing_2025} formalism was employed to parameterize the formation energies of symmetrically distinct arrangements of the refractory elements and vacancies on the bcc crystal structure. The eCE models accurately reproduced formation energies to within 8~meV/atom (\cref{fig:learning_curves}), demonstrating efficient parameterization across the high-dimensional composition space. We computed equilibrium vacancy concentrations in the nonary alloy using the coarse-graining framework of \cref{sec:stat_thermo} and Monte Carlo simulations informed by the eCE model. The data-efficient parameterization of the eCE model, combined with rigorous computation of non-dilute finite-temperature chemical potentials and the coarse-graining framework of \cref{sec:stat_thermo}, enabled a thorough exploration of finite-temperature equilibrium vacancy concentrations in concentrated alloys with \emph{ab-initio} accuracy. Finite-temperature calculations revealed that alloys composed solely of groups 5 and 6 elements exhibit extremely low vacancy concentrations (\cref{fig:vacancy_conc}), approximately $10^{-7}$ at $\approx 2000$ K. The addition of group 4 elements significantly increases vacancy concentrations, reaching nearly $10^{-4}$ at $\approx 1700$ K in the equiatomic Senkov alloy. The eCE model allows us to explore chemical trends in vacancy concentration, derive insights into relationships between ordering and vacancy concentration, and analyze the effects of approximations used previously in the literature.

\begin{figure}
    \includegraphics{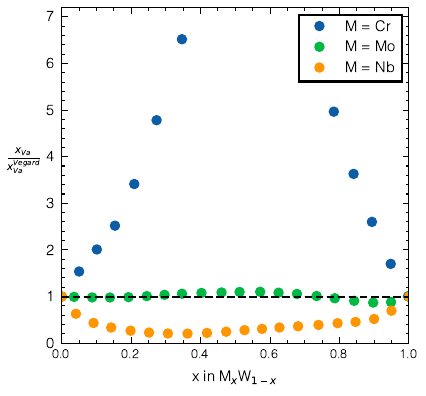}
    \caption{Deviation in vacancy concentrations from ideal solution behavior when Cr, Mo and Nb is alloyed with W at 2500K, as a function of the alloy composition. The black dashed line represents ideal solution behavior.}
    \label{fig:vacancy_conc_with_W}
\end{figure}

We begin by investigating the relationship between formation energies of binary orderings and vacancy concentrations in binary tungsten alloys. \Cref{fig:vacancy_conc_with_W} illustrates the variation in the ratio of actual to ideal vacancy concentrations across different alloy compositions for three prototypical tungsten binary systems. The ideal vacancy concentration ($x_{\textrm{Va}}^{\textrm{Vegard}}$) is computed by linearly interpolating between the vacancy formation energies of the pure elements and applying the Arrhenius relationship. We performed GCMC simulations at an elevated temperature of 2500~K to calculate ensemble averages of the vacancy concentration in all three alloys (following the approach of \cite{belak_effect_2015}). Monte Carlo simulations at such a high temperature were necessary to stabilize the disordered phase, especially in the phase-separating CrW alloy.

\Cref{fig:vacancy_conc_with_W} reveals three distinct trends in vacancy concentrations that correlate directly with the ordering energetics of \cref{fig:convex_hulls}. The Mo-W binary system exhibits nearly ideal mixing, as evidenced by the formation energies of symmetrically distinct atomic arrangements in \cref{fig:convex_hulls}. Orderings of molybdenum and tungsten on the bcc crystal structure have nearly identical formation energies regardless of the precise arrangement of atoms. Consequently, the vacancy concentrations in Mo-W alloys, represented by green circles in \cref{fig:vacancy_conc_with_W}, closely match the values predicted by ideal mixing. In contrast, Cr-W orderings demonstrate phase-separating behavior, indicated by their positive formation energies in \cref{fig:convex_hulls}. This phase separation tendency results in vacancy concentrations higher than those computed from a linear interpolation of the pure elements' vacancy energies. Finally, the Nb-W system forms several ordered ground states as shown in \cref{fig:convex_hulls}, and exhibits lower vacancy concentrations than would be expected in an ideal alloy.

The trends in vacancy concentration shown in \cref{fig:vacancy_conc_with_W} can be rationalized based on the bonding between elements. In the ideal Mo-W alloy, the elements are essentially interchangeable, causing vacancies to show no energetic preference for specific local coordination environments and thus producing nearly ideal vacancy concentrations. For the phase-separating Cr-W system, bonds between Cr and W atoms raise the crystal's energy and are thermodynamically unfavorable. Breaking these high-energy bonds becomes energetically preferable, resulting in increased vacancy concentrations. 
Conversely, Nb-W bonds are favorable, as arrangements of the two elements have negative formation energies (\cref{fig:convex_hulls}). As illustrated in \cref{fig:vacancy_conc_with_W}, breaking Nb-W bonds through the creation of a vacancy is unfavorable and causes a decrease in the vacancy concentration within Nb-W alloys compared to ideal mixing.

The correlation between ordering energetics (\cref{fig:convex_hulls}) and vacancy concentration (\cref{fig:vacancy_conc_with_W}) for binary alloys is found to persist even in concentrated alloys. As shown in \cref{fig:vacancy_conc}, substituting Mo with Ta in the equiatomic HfMoNbTiZr raises the equilibrium vacancy concentration. This is not expected as the vacancy formation energy of pure Mo is lower than that of tantalum. Mo forms several stable ordered phases with negative formation energies when alloyed with Hf, Zr, Ti, and Nb, as shown in \cref{SI:fig:convex_hull_group4,SI:fig:convex_hull_group5-6}. In contrast, alloying Ta with the elements of group 4 results in orderings with either positive or slightly negative formation energies (\cref{SI:fig:convex_hull_group4}). The bonds formed by Mo with group 4 elements are consequently stronger and more costly to break when a vacancy is formed. Vacancy formation in the Mo-containing alloy is therefore more difficult than in the Ta alloy, leading to lower vacancy concentrations.

Vacancy concentrations in refractory alloys can be substantially increased through the incorporation of group 4 elements into the alloy.
While adding Ti, Zr, and Hf has been traditionally considered beneficial for improving ductility in refractory alloys \cite{mak_ductility_2021}, these additions come with several drawbacks. For instance, increasing group 4 element concentration can degrade the high-temperature properties of the alloy. Higher concentrations of group 4 elements can also promote increased uptake of interstitial elements (oxygen, nitrogen, and carbon) \cite{gunda_understanding_2020}, potentially causing detrimental effects to mechanical properties. Excessive group 4 element concentrations could trigger undesirable martensitic phase transformations towards the hcp phase in these refractory materials \cite{natarajan_crystallography_2020}. Alloy designers typically carefully optimize group 4 element concentrations to attain sufficient ductility and high-temperature stability. The findings presented in \cref{fig:vacancy_conc} indicate that group 4 element concentrations should also be optimized to achieve appropriate vacancy concentrations to improve the processability of these materials.

\begin{figure}
\includegraphics{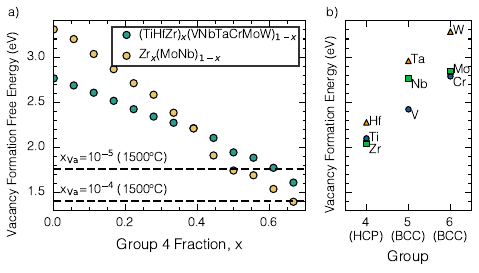}
\caption{a) Vacancy formation free energies in alloys with a varying amount of group 4 elements. The group 4 element fraction ($x$) is varied in a 9-component and ternary alloy. The vacancy formation energies required to achieve an equilibrium vacancy concentration of $10^{-4}$ and $10^{-5}$ at 1500°C are marked by the black dashed line on the plot. b) Vacancy formation energies in the pure elements.}
\label{fig:Ef_vs_group4}
\end{figure}

We investigate the variation of vacancy formation free energies with group 4 element composition in \cref{fig:Ef_vs_group4}. The composition of Ti, Zr, and Hf was varied from $x=0$ to 0.67 in a nonary (TiZrHf)$_x$(VNbTaCrMoW)$_{1-x}$ and ternary Zr$_x$(NbMo)$_{1-x}$ alloy. The ternary Zr-Nb-Mo system was selected as it is a prototypical alloy containing one element from groups 4, 5, and 6. We then compute the vacancy formation free energy from the equilibrium vacancy concentration through the Arrhenius relationship:
\begin{equation}
    \Delta G_f^{Va}=-k_BT \ln(x_{Va})
\end{equation}
where $\Delta G_{f}^{Va}$ is the vacancy formation free energy. To ensure that the vacancy formation free energies are accurate, we check that the respective alloys exist as a single phase using GCMC simulations and the dilute vacancy concentration approximation remains valid.

\Cref{fig:Ef_vs_group4}a illustrates the relationship between vacancy formation free energy and the concentration of group 4 elements. The calculations of \cref{fig:Ef_vs_group4}a reveal that $\Delta G_{f}^{Va}$ decreases with increasing concentration of group 4 elements. For alloys that rely on vacancy-mediated diffusion mechanisms, vacancy concentrations of approximately $10^{-4}-10^{-5}$ are generally necessary to achieve sufficient mass transport rates. \Cref{fig:Ef_vs_group4}a identifies the range of vacancy formation energies that would generate adequate vacancy concentrations at 1500\textdegree C. Both ternary and nonary alloy systems require $\approx 50-60\%$ concentration of group 4 elements to achieve these elevated vacancy concentrations. Interestingly, increasing the group 4 element concentration beyond 50\% is also found to stabilize the disordered bcc phase to lower temperatures. As suggested by the heat capacities of \cref{SI:fig:heat_capacity_2}, group 4 concentrations around $0.33$ increase the order-disorder phase transition temperature as compared with alloys that are either leaner or richer in group 4 element concentrations. 

The vacancy formation energy of the alloy decreases with increasing group 4 composition (\cref{fig:Ef_vs_group4}a). This trend is attributed to the low vacancy formation energies of the group 4 elements compared to those in groups 5 and 6 as shown in \cref{fig:Ef_vs_group4}b. The elemental vacancy formation energies were calculated using DFT in the hcp structure for group 4 and a bcc structure for elements in groups 5 and 6. The lower vacancy formation energy of group 4 elements correlates with their weaker interatomic bonding and lower melting points. Consequently, increasing the group 4 concentration lowers both the vacancy formation energy and the melting point of the multicomponent alloy. In addition to the intrinsically lower vacancy formation energies of group 4 elements, other effects such as charge transfer or atomic size mismatch could also play a role. For instance, as shown in \cref{SI:fig:solute_size}, the solute-vacancy binding energy generally becomes more negative with increasing atomic size mismatch, with large atoms like Zr and Hf exhibiting particularly strong binding. However, such qualitative trends are insufficient for accurately predicting vacancy concentrations in complex alloys. A rigorous statistical-mechanical treatment, like the one used in this study, is necessary to capture the strong interplay between local ordering, chemical bonding, and finite-temperature effects.

\begin{figure}
    \includegraphics{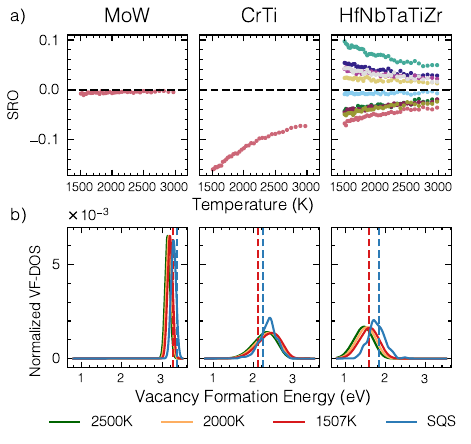}
    \caption{a) Plot of Warren-Cowley short-range order (SRO) parameters as a function of temperature for MoW, CrTi and HfNbTaTiZr. All SRO parameters are computed from nearest neighbor pairs of each unique pair of species in the system. b) VF-DOS computed for the three alloys at T=1507, 2000, 2500~K. The VF-DOS for a single SQS structure is also superimposed on the plots in blue. The vacancy formation free energies obtained from the SQS-VF-DOS and Monte Carlo simulations (1507~K) are indicated by the vertical dashed lines.}
    \label{fig:VF-DOS}
  \end{figure}

  The rigorous framework linking \emph{ab-initio} calculations to finite-temperature vacancy concentrations presented in this study enables us to assess other approximations used in the literature. Conventionally, vacancy concentrations are estimated based on vacancy formation energies computed within a supercell containing $\approx100$ atoms \cite{wilson_assessing_2023,nong_formation_2022,wang_first-principles_2022,xu_comparative_2022,lin-vines_defect_2022,roy_vacancy_2022,manzoor_factors_2021,zhang_statistical_2021,zhao_defect_2020,mizuno_defect_2019,li_first_2019,chen_vacancy_2018,zhao_effect_2018,zhao_defect_2016,zhang_ab_2015,piochaud_first-principles_2014}. In many studies, atomic arrangements of the elements are chosen to mimic the disordered state based on a special quasi-random structure (SQS) \cite{zunger_special_1990}. Exchange chemical potentials are also estimated from these finite-size supercells by performing Widom-type particle swaps and computing energy differences with DFT.

  \Cref{fig:VF-DOS} presents the density of states for vacancy formation energies (``VF-DOS'') \cite{zhang_ab_2022} and the corresponding short-range order (SRO) parameters for two binary alloys and the HfNbTaTiZr alloy. These systems were selected to assess the influence of SRO, chemical complexity, and supercell size on finite-temperature vacancy formation energies. To ensure direct comparison with the coarse-graining method described in \cref{sec:stat_thermo} and the finite-temperature results of \cref{sec:results_finite_temp}, all relevant energetics are computed using the same eCE model employed throughout this study.

The VF-DOS for each alloy is computed by evaluating vacancy formation energies from all single-species vacancy swaps across the Monte Carlo samples, using the true chemical potentials, and subsequently convoluting the results with Gaussian functions following the approach of \cite{zhang_ab_2022}. Additionally, we generate an SQS-based VF-DOS (SQS-VF-DOS) using a 128-atom ($4\times 4\times 4$ supercell of the conventional bcc structure) SQS cell. This SQS ordering is constructed to closely mimic the disordered state by optimizing correlations for the first three nearest-neighbor pairs, as well as first-neighbor triplet and quadruplet clusters. The SQS-VF-DOS is then computed by sequentially replacing each site with a vacancy and computing the vacancy formation energies with the excess chemical potentials estimated following the procedure described in \cite{zhang_ab_2022}. Widom-type particle swaps are performed on the SQS and the energy differences are computed with the eCE model. Finally, \cref{fig:VF-DOS}a shows the Warren–Cowley SRO parameters computed from the large-cell (2000 atoms) canonical MC simulations for the same alloys.

\Cref{fig:VF-DOS} reveals several differences in the vacancy formation energies, and spectra across all three alloys. The Mo-W alloy exhibits thermodynamically ideal behavior, at least with respect to nearest-neighbor SRO parameters. Even for the thermodynamically ideal alloy, the VF-DOS computed from rigorous Monte Carlo simulations differs from that obtained using a single SQS cell. Consequently, there is a small discrepancy between the vacancy formation energies calculated using the rigorous coarse-graining method (\cref{eq:approx_vacancy_concentration}) and the value derived directly from the SQS VF-DOS distribution. Thermodynamically non-ideal alloys like binary Cr-Ti and quinary HfNbTaTiZr show larger differences in the vacancy formation free energies calculated by these two methods. As illustrated in \cref{fig:VF-DOS}b, this discrepancy originates from significant differences in the shape and spread of the VF-DOS distributions obtained from MC simulations versus SQS calculations. These differing distributions lead to vacancy formation energies that differ by approximately $0.2-0.3$~eV for both Cr-Ti and HfNbTaTiZr. Consequently, the equilibrium vacancy concentrations predicted using the SQS are underestimated by a factor of 3 and 7 for Cr-Ti and HfNbTaTiZr. Interestingly, our calculations reveal a significant broadening of the VF-DOS in the Cr-Ti and quinary alloy as compared with the thermodynamically ideal Mo-W alloy in \cref{fig:VF-DOS}. The broadening of the VF-DOS correlates with the presence of SRO in the system, rather than the ``high-entropy effect'' as suggested by Zhang \emph{et al.} \cite{zhang_ab_2022}. 

Discrepancies in the vacancy formation free energies calculated from SQS and the rigorous coarse-graining method (\cref{sec:stat_thermo}) likely stem from differences in the constituent element chemical potentials within the SQS approach.  \Cref{SI:fig:mu_convergence} illustrates the differences in exchange chemical potentials as computed from Widom-type particle exchanges in a 128-atom cell and in a larger 2000-atom cell. The significant differences in these exchange chemical potentials may arise due to the smaller simulation cell size of the SQS cells. Employing chemical potentials derived from our finite-temperature Monte Carlo simulations to compute the SQS-VF-DOS yields closer agreement between the values of the vacancy formation energies computed from the two methods. Li and Schuler reached similar conclusions \cite{li_vacancy_2024,schuler_towards_2024} and proposed a correction methodology. 

The methodology outlined in this study rigorously accounts for the finite-temperature equilibrium distribution of chemical orderings in MPEAs when computing vacancy formation free energies. This is accomplished by directly sampling microstates from the equilibrium distributions in a sufficiently large simulation cell, with energies of these microstates being efficiently and accurately evaluated using the eCE model. In contrast, SQS-based approaches approximate the alloy as a perfectly random chemical decoration within a smaller supercell. This approximation may only be valid at very high temperatures or when the system remains close to ideal. Additionally, SQS-based approaches are only able to predict vacancy thermodynamics at the SQS compositions, while the eCE formalism utilizes energies from several compositions to interpolate vacancy interactions across the multicomponent alloy space. eCE models, when combined with MC sampling, enable the rigorous determination of vacancy concentrations without introducing any additional approximations across the entire composition space of a multicomponent alloy.

The coupling of eCE models with statistical mechanics enables the rigorous computation of vacancy formation free energies in multicomponent alloys containing a dilute concentration of vacancies. However, this methodology neglects contributions to the vacancy free energies arising from vibrational degrees of freedom. Vibrational entropy could play a significant role in the vacancy formation free energy of alloys containing a significant fraction of group 4 elements, as these elements are stabilized due to anharmonic contributions to the free energy. Additionally, at temperatures approaching the melting point of the solid, such vibrational free energies could be important even in alloys containing just elements from groups 5 and 6 of the periodic table. Furthermore, the formalism assumes the dilute vacancy limit and neglects the formation of divacancies or larger vacancy clusters.

\section{Conclusion}
  In this work, we demonstrated the advantages of using a data-efficient surrogate model, the embedded cluster expansion method (eCE), to connect electronic structure calculations with rigorous statistical mechanics methods for computing equilibrium vacancy thermodynamics in concentrated multicomponent alloys. We applied this method to a 9-component refractory multi-principal element alloy system containing group 4, 5, and 6 elements. Using the eCE model parameterized on first-principles calculations, we determined vacancy concentrations in specific alloy compositions through canonical Monte Carlo simulations. Our analysis of composition effects and local atomic interactions revealed that vacancy formation energies and equilibrium vacancy concentrations vary significantly with alloy chemistry. Group 4 elements enhance vacancy concentrations in multicomponent refractory alloys. We also find that formation energies of binary orderings can guide the search for alloying elements that significantly alter vacancy concentrations. Specifically, element pairs forming strong bonds decrease vacancy concentrations, while elements with unfavorable bonding increase vacancy concentrations. Solute-vacancy binding energy trends from dilute alloys persist even in complex concentrated alloys. This work provides alloy designers with a framework for identifying chemistries with target vacancy concentrations in high-dimensional composition spaces.

\section{Acknowledgements}
\label{sec:acknowledgements}
We acknowledge support from NCCR MARVEL, a National Centre of Competence in Research, funded by the Swiss National Science Foundation (grant number 205602) and the SNSF through project number 215178.

\bibliography{references}

\appendix
\section{Short-range order around a vacancy}
\label{sec:appendix_SRO_vacancy}

In addition to computing vacancy concentrations, it is useful to determine the local concentration of various chemical species around a vacancy. The average number of species $i$ that are in the vicinity of vacancies is given by:
\begin{equation}
  \label{eq:local_concentration}
  \langle N_{\alpha}^{i} \rangle = \sum_{\vec{\sigma}} N_{\alpha}^{i}(\vec{\sigma}) \frac{\exp(-\beta \Omega(\vec{\sigma}))}{Z}
\end{equation}
where $N_{\alpha}^{i}$ represents the number of atoms of type $i$ located within a coordination shell defined by $\alpha$ relative to the vacant sites.
In the dilute vacancy limit, \cref{eq:local_concentration} can be approximated as:
\begin{align}
  \label{eq:appendix_local_concentration_approx}
  \langle N_{\alpha}^{i} \rangle &\approx \frac{\sum_{\vec{\sigma}} \frac{\exp(-\beta \Omega(\vec{\sigma}))} {\tilde{Z}_{\textrm{alloy}}} \sum_{\vec{\nu}}N_{\alpha}^{i}(\vec{\nu})\exp(-\beta \Delta\Omega(\vec{\nu}))}{1+\xi}\nonumber\\
                                 &\approx \frac{\langle \sum_{\vec{\nu}}N_{\alpha}^{i}(\vec{\nu})\exp(-\beta \Delta\Omega(\vec{\nu}))\rangle_{\textrm{alloy}}}{1+\xi}\nonumber\\
                                 &\approx \frac{\theta_{\alpha}^{i}}{1+\xi}
\end{align}
where $\theta_{\alpha}^{i} = \langle \sum_{\vec{\nu}}N_{\alpha}^{i}(\vec{\nu})\exp(-\beta \Delta\Omega(\vec{\nu}))\rangle_{\textrm{alloy}}$
In \cref{eq:appendix_local_concentration_approx}, $\vec{\sigma}$ are alloy configurations without vacancies and $\vec{\nu}$ are orderings derived from $\vec{\sigma}$ with one of the sites containing species 1 being replaced by a vacancy. The local concentration of element $i$ within a coordination shell $\alpha$, $x_{\alpha}^{i}$, is computed as $x_{\alpha}^{i} = \frac{\langle N_{\alpha}^{i} \rangle}{\langle N_{Va} \rangle N_{\alpha}} = \frac{\theta_{\alpha}^{i}/\xi}{N_{\alpha}}$ where $N_{\alpha}$ is the number of sites in the coordination shell.

\section{Widom substitution method}
\label{sec:append-widom-subst}

We derive the Widom substitution method \cite{widom_topics_1963} for a binary alloy and generalize the expression to multicomponent alloys. The exchange chemical potential in a binary alloy containing $N_1$ atoms of specie 1 and $N_2$ atoms of specie 2 is defined as follows:
\begin{equation}
    \tilde{\mu}_2=\mu_2-\mu_1=G(N_1-1,N_2+1)-G(N_1,N_2)
\end{equation}
The exchange chemical potential is equivalent to the change in Gibbs free energy of a system in which one atom of species 1 is substituted with one atom of species 2. Expressed as a ratio of partition functions:
\begin{equation} \label{eqn:mu_tilde_partitions}
    \tilde{\mu}_2=-k_BT\ln\left(\frac{Z(N_1-1,N_2+1)}{Z(N_1,N_2)}\right)
\end{equation}
The partition functions $Z(N_1,N_2)$ and $Z(N_1-1,N_2+1)$ are given by:
\begin{equation}
    Z(N_1,N_2)=\sum_{\vec{\sigma}}
    e^{-\beta E({\vec{\sigma}})}
\end{equation}
\begin{equation}
    Z(N_1-1,N_2+1)=\frac{1}{N_2+1}
    \sum_{\vec{\sigma}}
    e^{-\beta E({\vec{\sigma}})}
    \sum_{\vec{p}(\vec{\sigma})}^{N_1}
    e^{-\beta \Delta E(\vec{p}(\vec{\sigma}))}
\end{equation}
in which $\vec{p}(\vec{\sigma})$ is the configuration obtained after exchanging an atom of species 1 with one atom of species 2 in configuration $\vec{\sigma}$ and the sum extends over all $N_1$ sites in $\vec{\sigma}$. The factor $1/(N_2+1)$ is required because the same configuration $\vec{\sigma}$ can be obtained from multiple configurations of $\vec{p}(\vec{\sigma})$, and we have to divide by $N_2+1$ to avoid overcounting configurations. Hence,
\begin{align}
    \label{eqn:final_partition_ratio}
    \frac{Z(N_1-1,N_2+1)}{Z(N_1,N_2)} =&     \frac{1}{N_2+1}
    \sum_{\vec{\sigma}}
    \frac{e^{-\beta E({\vec{\sigma}})}}{Z(N_1,N_2)}
    \sum_{\vec{p}(\vec{\sigma})}^{N_1}
    e^{-\beta \Delta E(\vec{p}(\vec{\sigma}))} \nonumber\\
    =& \frac{N_1}{N_2+1}
    \langle e^{-\beta \Delta E(\vec{p}(\vec{\sigma}))}
    \rangle_{\textrm{alloy},\textrm{crystal}}    
\end{align}

where $\langle e^{-\beta \Delta E(\vec{p}(\vec{\sigma}))} \rangle_{\textrm{alloy},\textrm{crystal}}$ is the average value of $e^{-\beta \Delta E(\vec{p}(\vec{\sigma}))}$ computed over the alloy and crystal.
Substituting \cref{eqn:final_partition_ratio} into \cref{eqn:mu_tilde_partitions}, we obtain
\begin{equation}
    \tilde{\mu}_2=-k_BT\ln{\frac{N_1}{N_2+1}}
    -k_BT\ln{\langle e^{-\beta \Delta E(\vec{p}(\vec{\sigma}))}
    \rangle_{\textrm{alloy},\textrm{crystal}}}
\end{equation}
In the thermodynamic limit, the first term represents the ideal exchange chemical potential while the second term represents the excess exchange chemical potential as described in other studies \cite{sindzingre_partial_1987, sindzingre_calculation_1989}. In a simulation, the exchange chemical potential $\tilde{\mu}_i$ can be estimated by sequentially swapping each atom of species 1 with species $i$ across all its sites and configurations sampled in the Monte Carlo trajectory, and computing the associated energy change for each exchange.

\end{document}